\documentclass{aa}
\usepackage{amssymb}
\usepackage{graphicx}
\begin{document}

%
   \title{The Carlsberg Meridian Telescope CCD Drift Scan Survey\thanks{
   The catalogue is available at the CDS via anonymous ftp to
cdsarc.u-strasbg.fr (130.79.128.5)
or via \hfill\break
http://cdsweb.u-strasbg.fr/cgi-bin/qcat?J/A+A/
}}

\titlerunning{The CMT CCD Drift Scan Survey}


   \author{D. W. Evans\inst{1}
           \and
	   M. J. Irwin\inst{1}
	   \and
	   L. Helmer\inst{2}
          }

   \offprints{D. W. Evans,
   \email{dwe@ast.cam.ac.uk}}

   \institute{Institute of Astronomy, Madingley Road, Cambridge, CB3 0HA, UK
         \and
             Copenhagen University Observatory, Juliane Maries Vej 30,
             DK-2100 Copenhagen {\O}, Denmark
             }

   \date{Received / Accepted }

   \abstract{

This paper contains the general data reduction methods used in processing
the data from the Carlsberg Meridian Telescope CCD Drift Scan Survey. An
efficient method to calibrate the fluctuations in the positions of the
images caused by atmospheric turbulence is described. The external accuracy
achieved is 36 mas in right ascension and declination. A description of the
recently released catalogue is given.

   }

   \maketitle

%

\section{Introduction}

The Carlsberg Meridian Telescope (CMT) has recently undergone a major upgrade.
A
2k by 2k CCD camera has been installed with a Sloan $r'$ filter operating in a
drift scan mode. With the new system, 
the effective exposure time is about 90s,
the magnitude limit is $r'_{\rm CMT}=$17 and the
initial positional accuracy is in the range 0.05$''$ to 0.10$''$.

The main task of the CMT is to map the sky in the declination range
$-$3$\degr$ to $+$30$\degr$ with the aim of 
providing an astrometric, and
photometric,
catalogue that can accurately transfer the Hipparcos/Tycho reference frame
to Schmidt plates. 
A secondary survey is also planned that extends the declination range
covered to $-$15$\degr$ in the South and $+$50$\degr$ in the North.
Projects similar to the CMT (UCAC, \cite{Zacharias} and 
CMASF, \cite{CMASF}), which 
include the 
Southern
hemisphere, will also be able to provide astrometric calibration for VISTA
and other deep wide-field surveys.

The two main systematic errors affecting the data are caused by image
motions due to long timescale atmospheric turbulence
and charge transfer efficiency (CTE) problems linked to the CCD.
Methods are described on how to calibrate these errors.


\section{The telescope system}

The telescope is a Grubb Parsons refractor with an objective of 178 mm
diameter and focal length 2.66 m. After initially being located at
Brorfelde, Denmark, the telescope was moved in 1984 to La Palma in the Canary
Islands to take advantage of the better observing conditions there.

Originally, the detector used was a scanning-slit photoelectric micrometer,
but was replaced in 1998 by a CCD camera operating in a drift scan mode.
This was a major change in the method of observing, since relative
astrometry with respect to a dense grid of standards within the same
data frames would be used rather than absolute astrometry and offsetting the
telescope with respect to the standards. 

The two significant advantages of a CCD system are that fainter stars can be
observed and that many stars can be observed simultaneously. This has
increased the number of stars that can be observed in a night by a factor of
more than 100.
However, there is a disadvantage with the new system in that close to the
celestial pole the images become distorted. This is discussed further in
Section~\ref{fieldcorrections}. Although this restricts normal
observing to declinations South of $+60\degr$, it is more than compensated by
the amount of high-quality data that it produces.

More recently (April 1999) the CCD system was upgraded to a larger detector
(Kodak 2k$\times$2k with 9$\mu$m pixels) and a filter equivalent to the
Sloan Digital Sky Survey $r'$ passband was fitted.

The increase in the CCD size has a number of advantages which include:
completing the survey faster due to a larger field of view; providing a
deeper survey due to longer exposures; and improvements in the calibration
due to increased frame sizes. Also, with the fitting of the $r'$ filter, the
project is now also able to provide photometry on a commonly-used
photometric system.

The new CCD pixel size corresponds to $0.7''$ and considering that
the median seeing at the telescope is just under $3''$ the images
are well sampled. It should be noted that the site seeing is much better
than this.

The CCD can
be cooled to $-65\degr$C by a Peltier cooler. Currently, the chip is cooled
to $-30\degr$C since this reduces the effect of a charge transfer
efficiency (CTE) problem with the chip (see Appendix~\ref{secCTE}). The
higher operating temperature does not affect our magnitude limit.

These improvements with the CCD and the new filter have increased the number
of stars observed by about 4 times as many stars per night than with the old
CCD system. The current magnitude limit is $r'_{\rm CMT}=17$ and between 100,000 and
200,000 stars a night are observed. On a typical night, more than 50 square
degrees are covered.

More details about the telescope can be found in \cite{vistas} and about the
recent upgrades in \cite{tech}. A summary is given in Table~\ref{tableparam}.

\begin{table}
\caption{A summary of the telescope and camera parameters for the current 
configuration.}
\centering
  \begin{tabular}{ll}
    \hline
    Telescope: & Located on La Palma, Canary Islands\\
               & 178 mm objective\\
	       & 2.66 m focal length\\
    \hline
    Camera:    & CCD chip -- Kodak (KAF-4202 Grade:C1)\\
               & 2060$\times$2048 pixels\\ 
	       & Pixel size 9$\mu$m (0.7$''$)\\
	       & CUO built\\
	       & Operating temperature $-$30$\,\degr$C\\
    \hline
    System:    & Automatic and remotely controlled\\
               & Drift scans\\
	       & \qquad (generates $\sim$3 Gb of data per night)\\
	       & Data automatically parameterized\\
	       & \qquad (reduced to 6-7 Mb)\\
	       & Daily reductions take about 30 min.\\
	       & 100,000--200,000 stars observed per night\\
	       & Calibrated with respect to Tycho 2\\
	       & Magnitude limit (Sloan) $r'_{\rm CMT}=$17\\
    \hline
  \end{tabular}
\label{tableparam}
\end{table}


\section{Observing strategy}

The effective width of the CCD is 2060 pixels which corresponds to drift
scans of width $23'$. The selection of where to observe is constrained by
the survey nature of the current project. Since specific objects are not the
targets, the observing is carried out at evenly spaced declinations (every
$0.25\degr$). This allows for a uniform overlap of $8\,'$ in declination
between scans, which provides enough data to quantify the difference between
two adjacent frames and calibrate the atmospheric fluctuations (see
Section~\ref{FlucSec}).

Each day preliminary calibrations are carried out which provide quality
control information which is then fed into the observation selection
programme. The basic principle of the selection programme is to maximize the
lengths of the observations. The reason for doing this is to minimize the
number of intervals between observations during which the telescope is
moving to a new declination and the CCD is being read out and thus maximize
the amount of time collecting data.

Various declination zones have different priorities, so that the primary
survey area ($-$3$\degr$ to $+$30$\degr$ declination)
is completed before other areas are observed.
Additionally, a $20^\circ$ zone of
avoidance around the Moon is used. There is also a minimum observation
length of 20 minutes, otherwise the Tycho 2 preliminary calibration will
probably fail due to too few standards being available. A typical drift scan
lasts about an hour, although exposures up to 5 hours have been made.

Additional calibrations are carried out off-line at Cambridge in order to
account for the CTE and atmospheric fluctuation problems. From these it is
possible to identify further problem data frames that need to be
reobserved. This information is then fed back to the selection programme after
the calibrations have been visually inspected.


\section{Astrometric data reductions}

In this section the main astrometric data reduction is discussed. The data
from the CCD is reduced automatically as soon as the observation
has been completed. 
The data reduction system automatically detects and parameterizes images and
produces other data monitoring diagnostics.
During the day, the observer then carries out a
preliminary calibration which serves both as a first pass calibration and
also as a quality control check. Most of the analysis for systematic effects
is carried out on the data at this stage. Finally, the data is accumulated
and a catalogue formed.


\subsection{Image Analysis}\label{ImageAnalysis}

The data processing pipeline is designed to ingest the variable-length drift 
scan two-dimensional images and automatically detect and parameterize objects
located on the frames.  The nature of the remote operation precluded long term 
storage of the drift scan images ($\sim$3 Gbytes/night) and correspondingly
defined the automated nature of the processing pipeline.  By storing only
relevant information (position, intensity, shape) for each detected object, 
a factor of $\sim$100 compression over the raw data results, with virtually no 
loss of relevant information.  The resulting object catalogues then form the
basis of all subsequent reductions.  Generic data quality control information 
is also extracted from the pipeline products and contributes to monitoring the
health of the overall system.

As usual, the first part of the data processing involves removing the
instrumental signature, which in this case reduces to a one-dimensional
correction perpendicular to the drift scan direction.  In theory, with a
modern CCD camera, these corrections involve correcting for the DC bias level
and then flatfielding out the remaining systematic effects.  However, in
practice, we found that the additive bias correction was not simply a
constant level across the frame and indeed was difficult to disentangle from
the effects of the multiplicative flatfield correction.  After a series of
on-sky tests (see Appendix~\ref{FlatFielding}) we found that the dominant
contribution to background variations across the scan direction was additive
in nature and that after correcting for this no significant flatfield
variations (ie. $\gtrsim$1\%) remained.

Consequently, the first pass preprocessing consists of using the underscan and 
overscan regions to monitor the overall bias (or zero) level of the device, 
while the active part of the CCD defines the differential additive correction
to be applied in subsequent processing stages.  We suspect the control and
readout electronics introduce the varying (but repeatable over intervals
$\approx$nightly) bias level across the CCD rows during clocking out each row 
of data.  The bias correction is defined as the median of the data in each
column with respect to the median level of the underscan and overscan regions.

In the same pass through the data, the general one-dimensional background 
variation down the scan direction (ie. as a function of RA/time) is also 
recorded, again using the median level, with an effective scale length of $\sim$
1 arcmin.  At the same time as the background variation is monitored, a robust
estimate of the {\it rms} sky noise is made based on the Median of the 
Absolute Deviations from the median (MAD estimator - see \cite{MAD}
for more details).  Finally, an overall estimate of the sky 
noise level for the whole frame is made from the median of the MAD estimates.

In normal conditions, the sky is sufficiently uniform over the cross-scan 
direction and at such a low level ($\approx$10 counts cf. readout noise of 7 
counts) that tracking its variation in the scan direction is sufficient.
All bias and sky level estimates are saved for subsequent diagnostic and
data quality control checks.  

On the second, and final, pass the background-corrected data is then searched 
for discrete astronomical objects using the techniques described by Irwin 
(1985, 1996).  Briefly, this consists of using a matched isophotal detection 
algorithm to locate regions of connected pixels above a definable detection 
threshold (typically 1.5 $\times$ sky noise level).  Each such region defines 
a potential astronomical object, which may be single or multiple.  The 
contiguous pixel lists are then searched for the presence, or otherwise, of 
multiple components.  The flux is appropriately partitioned, if needed, and 
various image parameters describing the location, intensity and shape are 
computed (see \cite{Irwin1996}).  

Since the primary driver of the project is the astrometric precision 
attainable, the choice of image parameterization method was dominated by this
consideration coupled with the requirement for the reduction to be fast and
completely automatic.  

Precision astrometry (usually) depends on minimizing both systematic and
random errors, and in drift scanning with the CMT system, as we discuss
later, the systematic errors can be at least as large as the random errors
even for the fainter images.  An additional problem with the CMT in drift
scan mode (and in most other imaging systems) is that the Point Spread
Function (PSF), in general, varies over the frame in both the cross-scan and
scan directions making it extremely difficult to attain the theoretical
error bounds for the random part of the error and might also introduce a
further systematic source of error. For the CMT we have therefore adopted a
modified intensity-weighted centre-of-gravity (CoG) method as a compromise
between robustness, ease of implementation and accuracy attainable.  We have
found that it is possible to design a simple weighted CoG method that
approaches the accuracy achievable with $\bf perfect$ 
PSF fitting without computing the PSF or invoking a non-linear iterative
scheme.

For example, it is well known (eg. \cite{Irwin1985}) that for bright images 
dominated by Poisson statistics, an intensity-weighted centre-of-gravity is 
the optimum estimator for location, whereas for faint images dominated by a 
constant Gaussian error, unweighted PSF fitting is 
optimal.  For the CMT in drift scan mode, noise due to the sky background is 
generally much smaller than the pixel readout noise, so to a very good 
approximation all pixels see a constant Gaussian noise with photon noise from 
the object pixels added in quadrature. 
\cite{Irwin1985} demonstrated that the Maximum Likelihood solution to 
this problem could be thought of as a modified CoG method where the optimal
additional weighting depends on the noise properties, the PSF and iteratively 
improving the estimate for the centre of the image.  The ideal extra weighting
function is essentially a smooth version of the original image, centred on
the (unknown) location where the shape of the smoothing function depends on 
the local signal-to-noise.  In our algorithm, we simply fix the smoothing 
function to be the same as used in the detection filter stage, since this
is already available, and use this smooth image (relative to local sky) to 
define the extra weighting to use in the CoG method.  Since the detection 
filter PSF is symmetric the extra weighting is automatically centred on the
(unknown) image position and therefore requires only a single pass through
the data.  The improvement over standard CoG methods is best for faint 
images (as expected) but does not noticeably degrade the performance for 
bright images.

The single pass nature of the method means that, if needed, this algorithm is 
fast enough to process the data in real time using only a modestly resourced
PC.  It is also worth emphasizing that the image analysis described above is 
completely automatic in nature and is run on each drift scan frame as soon as 
it has finished being taken.  The processing is invoked by an automatic data 
monitoring script that has overall control of the image analysis.  Roughly two
weeks after being taken the raw data is deleted due to lack of suitable 
on-line disk space and operational constraints of running the telescope 
remotely.  This gives sufficient time to track down and analyse gross system 
faults via FTP transfer of selected full data frames.


\subsection{Initial astrometric fit}\label{findact}

The main principle behind the measurements made by the CCD system is the use
of relative astrometry. Although the design of the original telescope
(\cite{vistas}) was with absolute astrometry in mind, better accuracy can
be achieved by calibrating with respect to standards within the same
data frames rather than relying on the accuracy of the telescope itself.

The astrometric standards used are those of Tycho 2 (\cite{tycho}).
The mean star density of this catalogue ranges from
25 to 150 stars deg$^{-2}$ and has a magnitude limit of V$\sim$11.5. 
Not all the stars in Tycho 2 are suitable for use as standards for this
project. Entries have been excluded if they have no proper motion data, if
they are double stars 
or have poor
astrometric solutions. This excludes about 30\% of the entries in Tycho 2.
In doing so, we have erred on the side of caution in order to improve the
robustness of the calibration.
For the drift scans used in this project this corresponds to a standard star
density of between 7 and 40 stars per degree of scanning (4 minutes) on the
equator.

During the analysis of the data, CCD images are excluded if they are too near
the edges or the terminal ramps, since these images would be
distorted and have poor astrometry. Images are flagged if they are
elliptical and do not take part in the calibration, however, they are included
in the final catalogue. For the magnitude range of this catalogue an
elliptical image is usually indicative of a double/multiple image rather 
than a galaxy.
Saturated images are also flagged. The level at which saturation occurs is
60,000 counts and corresponds to approximately an $r'_{\rm CMT}$ magnitude of 8--9.

Various corrections are carried out to the data before the fitting to the
standards is attempted. This is in order to improve robustness and to remove
certain systematic errors that would not be removed by later calibrations.
One set of corrections are the field corrections (see
Section~\ref{fieldcorrections}). These are repeatable systematic correction
needed as a function of declination. Since the CCDs are read out in drift
scan mode, no corrections as a function of right ascension 
are necessary. The
positions of the CCD images are also converted from apparent to mean
positions. Even though relative astrometry is needed the frames are
sufficiently long that non-linear terms affect the matching and must be
allowed for.

The matching of the Tycho 2 stars to the CCD images was done in a two-pass
process in order to improve robustness. In the first pass only the brightest
stars were used. This limits the chances of mismatches occurring in the event
of the telescope having a small positioning error. The initial search radius
was 200 pixels ($140''$). Following the matching a 4-parameter model (scales
and offsets) was fit to the data with an iterative 3-sigma cut to reject 
outliers.

After this initial fit was carried out, all CCD positions were transformed
using this 4-parameter solution and a second match was performed between all
CCD images and Tycho 2 standards using a smaller search radius
(20 pixels, $14''$).
This time the solution used a general 6-parameter linear fit, 
again with an iterative 3-sigma cut. The form of this solution was

\begin{equation}
\begin{array}{ll}
x_\mathrm{new}&=ax_\mathrm{old}+by_\mathrm{old}+c,\\
y_\mathrm{new}&=dy_\mathrm{old}+ex_\mathrm{old}+f.\\
\end{array}\label{Eq6Param}
\end{equation}

The CCD positions were then transformed for a final time and converted into
right ascensions and declinations.

During this initial calibration phase various statistics are accumulated and
output as diagnostics. Examples of these are the solution standard
deviations, the magnitude limit (see Section~\ref{PhotSec}) and average
image shape. The standard deviations give a clear indication of the
magnitude of the image fluctuations (see Section~\ref{FlucSec}) while the
average image widths give an estimate of the seeing conditions. These two
are weakly correlated. 
The average image ellipticity gives a good
indication that the drift scanning rate is correct.
All these diagnostics act as a
primary quality control which then feeds back into the selection programme.


\subsection{Calibration of Fluctuations/Image Motion}\label{FlucSec}

One of the main problems with drift scan surveys is the astrometric
fluctuations, caused by atmospheric seeing effects (\cite{Hoeg};
\cite{Benevides-Soares}), 
which typically have a wavelength of a couple of minutes
and a typical  peak-to-peak amplitude of a few tenths of an arc second.
These fluctuations cause systematic
errors in RA and declination as a function of RA. Even with the Tycho 2
catalogue, not enough standards are present to calibrate these fluctuations
directly. 

The main approach used by other groups so far has been to use a subcatalogue
of positions formed from repeat observations 
eg.~Bordeaux (\cite{Viateau}). Each area is observed on a
number of nights and, after a simple fit is applied to Tycho 2, the positions
get added to a subcatalogue. Any nights that seem to be much worse then
any others are rejected. For each star an average position is formed which
should reduce the effect of the nightly fluctuations. This makes the
reasonable assumption that
the fluctuations are not correlated from one night to the next.
If a
position is required for a particular night, eg.~for a planet, the
subcatalogue is used to calibrate the fluctuations and produce a position
for that night. The problem with this technique is that it considerably
increases the amount of time required to cover the sky. Also, even with a
large number ($N$) of observations, the reduction
in the amount of fluctuations left in the subcatalogue will only be
by a factor $\sqrt{N}$.

The technique used by the CMT is to use the Tycho 2 stars to calibrate out the
fluctuations and get around the problem of sparsity of standards by using
overlapping frames.

\begin{figure*}
\resizebox{\hsize}{!}{\includegraphics[angle=270]{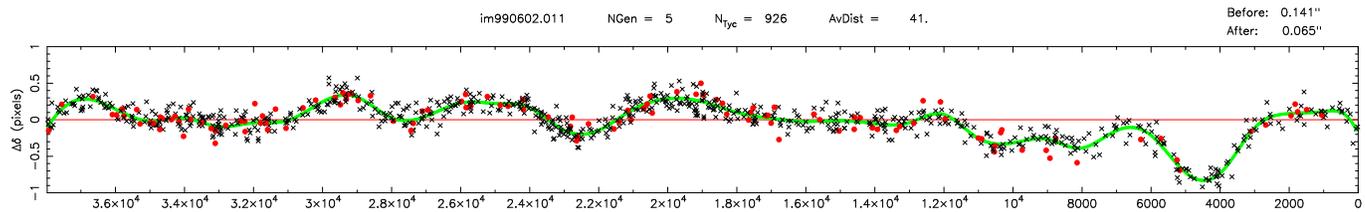}}
\caption{This shows an example of the Tycho 2
calibration functions for declination
for a particular frame. The red points are the residuals
from the primary frame. The crosses show the
data from the overlapping frames.
The green line is the derived final calibration
function. The units of the residuals (CMT $-$ Tycho 2)
are in pixels ($\sim$0.7$''$).
In the calibration programme the RA values are normalized in order to
improve the numerical
stability of the solution. The actual range is 38,500 pixels and
is the same as in Figure~\ref{Fig3}.}
\label{Fig2}
\end{figure*}

\begin{figure*}
\resizebox{\hsize}{!}{\includegraphics{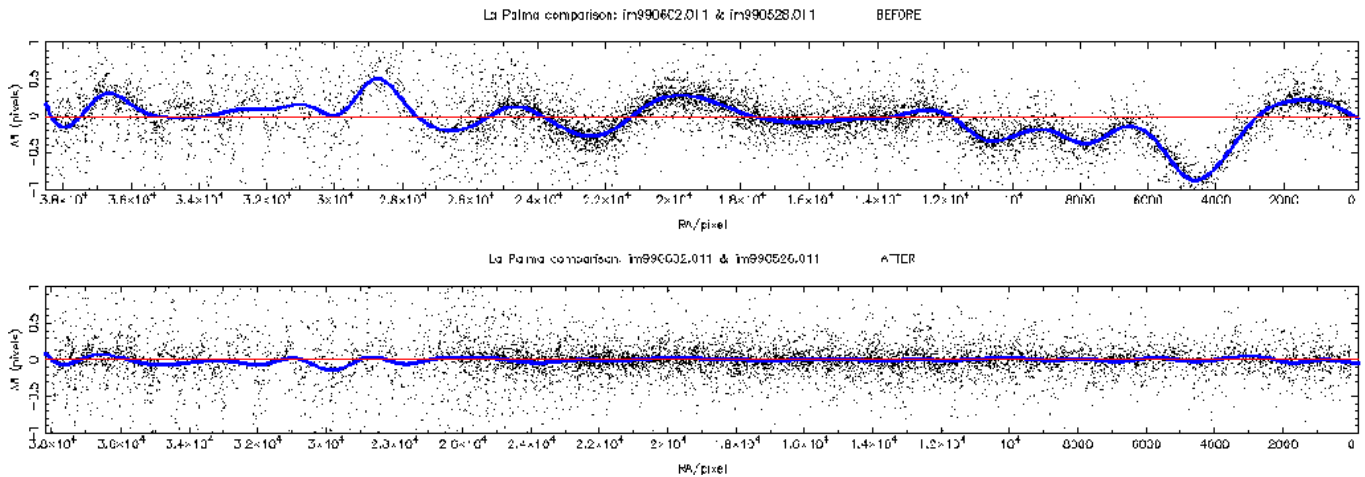}}
\caption{This shows an example of a night-to-night comparison for declination.
The top plot is before the calibrations have been applied and the bottom one
afterwards.
Most of the fluctuations come from one night (2nd June 1999) as can be seen
from the calibration plot (Figure~\ref{Fig2}). The reduced number of stars
and increased number of outliers on the left of the plots is due to cloud
coming over during the exposure. While this affects these plots, which
mainly show the fainter stars, the calibration plots only contain bright
stars
($r'_{\rm CMT}<$12)
and are thus not affected.}
\label{Fig3}
\end{figure*}

Observations of the survey are carried out
on a declination grid of 15 arc minutes. Observations taken on
different nights at adjacent declinations will 
have an overlap of about 36\%. This is
sufficient to define a transfer function (using all the stars in the
frame) which characterizes the difference
in the fluctuations between two nights. By applying these differences
to the positions in the second frame the fluctuations of the secondary night 
are effectively transformed
into those of the primary.

The transfer function is simply a set of cubic splines. The initial set of
knots are placed at intervals of 1000 pixels ($\equiv$ 45 seconds of time).
If a smaller interval
was used, then the transfer function would
map out a higher frequency than would be valid. This is determined by the
highest frequency of fluctuations observed, which is in turn determined by
the effective exposure time ($\equiv$ 2000 pixels $\equiv$ 90s).
After the initial placement of knots has been made, the data is checked to
see if there are enough points (10) present to define the splines reliably.
If this is not the case, then the two knots in question are merged. This is
carried out iteratively until the criterion is met for all knots.

The transformed positions from the secondary nights are then added to those
from the primary night. Further frames that overlap this ``new'' extended
frame are searched for and added in a similar way. This process can be
carried on until there are a sufficient number of standards in the frame.
However, the further away from the original data the transfer functions are
calculated, the greater the problems arising from the propagation of errors
from multiple application of transfer functions.

Investigations were carried out to determine the optimal number of
generations to go from the original data. This was done using the residuals
with respect to the Tycho 2 standards within the original frame. It was
found that the best value varied between 4 and 9 generations, but the higher
values tended to be less robust. A value of 5 generations was chosen as a
reasonable compromise which produces consistently good results.

After building up a large frame, Tycho 2 stars are matched and the
fluctuations mapped using a calibration function. In this case, the
calibration function characterizes the difference between the primary night
and the Tycho 2 positions. This is then applied to the positions of the
primary night (only) to produce a calibrated output file. While transformed
positions from the overlapping frames could be output at this point it is
not done since their accuracy has been degraded via the night-to-night
transfer functions. It is better to run the algorithm separately for each
frame.

A similar procedure is carried out to calculate the calibration function to
that used in determining the transfer function. A density of higher than 1
Tycho 2 star per 200 pixels is needed in order to obtain a reliable final
calibration function.

Figure~\ref{Fig2} shows an example of the final Tycho 2 calibration function
for 5 generations of overlapping frames. This frame was chosen 
since it was from a particularly bad night and has very large fluctuations.
Note that the rightmost trough of the declination plot does not have enough
primary standards to define it adequately, but the addition of the standards
from the overlapping frames provides enough information to calibrate the
trough correctly.

After calibration, the residuals with respect to Tycho 2 are normally between
50 and 80 mas and mainly reflect errors in Tycho 2. Because of this, it is
difficult to estimate the CMT external errors from these residuals.

After applying this calibration method to two overlapping 
frames a comparison was
carried out. The results are shown in Figure~\ref{Fig3}. The two frames
are from the nights of the 28th May and 2nd June of 1999.
Most of the differences seen originate from the fluctuations on the
2nd June (see Figure~\ref{Fig2}).
As can be seen in the AFTER plot, most
of the fluctuations have been removed. 


\subsection{Other systematic effects}\label{fieldcorrections}

\begin{figure}
\resizebox{\hsize}{!}{\includegraphics[angle=270]{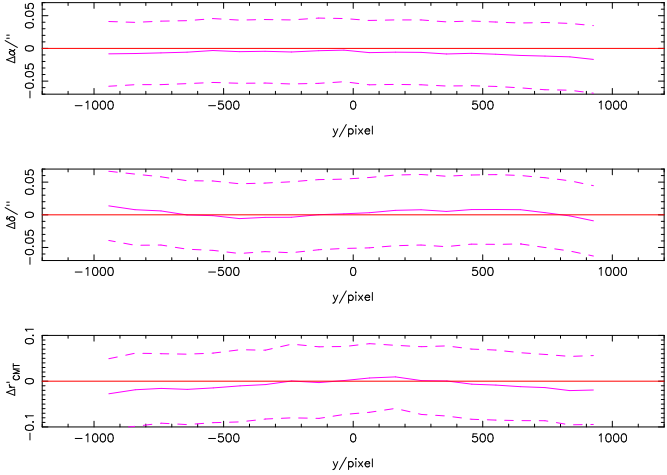}}
\caption{These plots show the accumulated residuals with respect to Tycho 2
as a function of $y$ ($\equiv$~declination) for stars brighter
than $r'_{\rm CMT}=10$. The solid line shows the median
of the distribution and the dotted line the width.}
\label{field}
\end{figure}

Each time the preliminary calibration programme is run, the residuals with
respect to Tycho 2 in right ascension, declination and magnitude are
accumulated. The systematics in this data are shown in Figure~\ref{field}.

Although these systematic errors are not very large in comparison with the
random errors for bright objects, they should be
corrected. Cubic polynomials were fitted to the data and applied in
the calibration. 

Investigations were carried out to see if the positional distortions were
comparable with known effects. The effect of tangential projection can
easily be estimated using a simplistic ``$\tan\theta$'' analysis. This is
because drift scanning has smeared out most of the projection effects in
right ascension and simplified the problem to a one-dimensional one. This
will cause a distortion in declination of order of 3~mas. The systematic
effect observed in declination is around 20~mas peak-to-peak.

As mentioned above, one of the effects of drift scanning is to smear out the
projection effects (\cite{GibsonHickson}; \cite{Stone}). Using the equations
below, taken from \cite{Taff}, the effect of image distortion can be
calculated from projecting a spherical surface onto a flat plane (CCD chip).

\begin{eqnarray}
\xi  &=&{\cot\delta\sin(\alpha-\alpha^*)} \over 
        {\sin\delta^*+\cot\delta\cos\delta^*\cos(\alpha-\alpha^*)}\\
\eta &=&{\cos\delta^*-\cot\delta\sin\delta^*\cos(\alpha-\alpha^*)} \over 
        {\sin\delta^*+\cot\delta\cos\delta^*\cos(\alpha-\alpha^*)}
\end{eqnarray}

(For further detail regarding these equations see \cite{Taff}).

Not only was the width of the image distortion investigated, but also the
median position of the image. The width analysis produces results very
similar to those of Figure~10 in \cite{Stone}. 

For the median of the positions, in right ascension, no systematic shift was
observed since the effect of drift scanning is to distort the image
symmetrically as long as the image is exposed equally either side of the
meridian, ie.~a complete drift scan over the CCD chip which is centred on
the meridian. In declination, the situation is different in that the image
is distorted towards the celestial pole. However, in order to affect the
astrometry it is the difference in this distortion between the top and
bottom of the CCD chip that is important. Results from the analysis showed
that this effect is very small ($\sim$4 mas) and not greatly affected by the
declination of the observation (for declinations less than $60^\circ$).
Also, when a fit is carried out with respect
to the Tycho 2 standards, the solution of the declination scale removes most
of this systematic error.

The conclusion of this analysis is that projection effects and differential
image distortion does not account for the systematic effects observed in
right ascension and declination.

The systematic effect in magnitude as a function of $y$ (equivalent to
declination) is discussed in Appendix~\ref{FlatFielding}.

A small systematic effect also exists in declination as a function of
colour. This is caused by the wavelength dependence of atmospheric
refraction. Using a spectral atlas it is possible to calculate the
correction appropriate for the passband defined by the $r'$ filter and the 
response of the KAF-4202 CCD chip. This method is outlined in
\cite{Paper2}, except that for the calculations in this paper the data from
\cite{Pickles} was used rather than from \cite{GS} and an average
atmospheric pressure of 780 mbar was used.

\begin{figure}
\resizebox{\hsize}{!}{\includegraphics[angle=270]{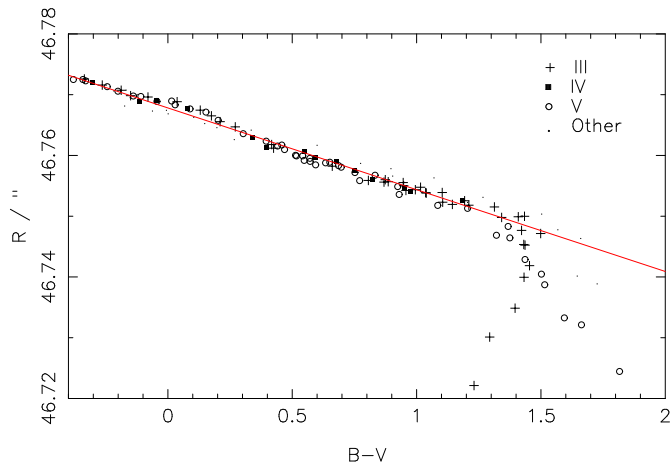}}
\caption{This plot shows the constant of refraction as a function of
(B$-$V)$_{\rm J}$. The various symbols stand for different types of star.
The line drawn is the best fit line to the points with (B$-$V$)_{\rm J}<$1.2
and is equivalent to that given in Equation~\ref{act}.}
\label{refractplot}
\end{figure}

These corrections are within 1 or 2 mas to those given in Table 2 of
\cite{Stone1} after accounting for the difference in atmospheric pressure. 
However, there is a larger difference for the
reddest stars (B$-$V$>$1.4). This is possibly due to detailed differences in
the passbands of filters used at the two telescopes. The correction that
should be applied to the data, $\Delta R$, can be calculated from
Equation~\ref{act}. This relation is also shown in Figure~\ref{refractplot},
which shows the detailed results from the spectral flux analysis. From this
diagram can also be seen that for the reddest stars the relation deviates
from linear and might provide another reason for the slight difference with
the results in \cite{Stone1}.

\begin{equation}
\Delta R = -13.5 [{\rm (B-V)_J}-0.60]  \enskip({\rm mas})
\label{act}
\end{equation}

Although the analysis calculates an absolute constant of refraction, it is
only a relative term that is required as a correction since the calibration
with respect to Tycho 2 has already accounted for the average refraction
term. In the above equation, the offset used is the average colour of
the Tycho 2 stars used in the calibrations. This was found to be
$\rm (B-V)_J=0.60$ (equivalent to $\rm (B-V)_T=0.71$, see Equation~1.3.20 of
Volume 1 of \cite{hippar}). This colour corresponds to that of a G0 star.

This correction can then be applied to the declination using the following
formula:

\begin{equation}
\delta_{\rm corr}=\delta+\Delta R\tan z/3.6\times10^6 \enskip({\rm degrees}),
\end{equation}

where $\Delta R$ comes from Equation~\ref{act}
and $z$ is the zenith distance, with this being positive for stars North
of the zenith. Since colours are generally not available for the stars in
the survey, this correction has not been applied, however for the majority
of stars in the survey ($\rm 0.0\lesssim(B-V)_J\lesssim1.5$),
$\Delta R$ varies by about $\pm10$~mas and consequently the correction to
declination will be less than 10 mas for all the survey
cf. the astrometric accuracy for the bright end of 36 mas.


\section{Photometric data reductions}\label{PhotSec}
The main part of the photometric reductions are carried out by the same
calibration programme described in Section~\ref{findact}. The photometric
data used as the standards are the $B_T$ and $V_T$ values from the Tycho 2
catalogue. Although there are catalogues with higher accuracies at the faint
end, Tycho 2 is uniquely homogeneous and dense. However, not all Tycho 2
stars were used. In order to make the photometric reductions more robust,
stars identified as variable were excluded from the calibration. Due to the
nature of the reduction process used to create Tycho 2, variability
information is not available in that catalogue. For this, the original Tycho
data (\cite{hippar}) must be used.

The intensities determined from the images (see Section~\ref{ImageAnalysis})
are first converted into magnitudes ($m=-2.5\log_{10}i$) and are then 
corrected in order to take into account
the difference between isophotal and total magnitudes (see
Section~\ref{APMCAL}). Following this, the calibration then simply consists 
of determining the zero point of the magnitude scale. 

\begin{figure}
\resizebox{\hsize}{!}{\includegraphics[angle=270]{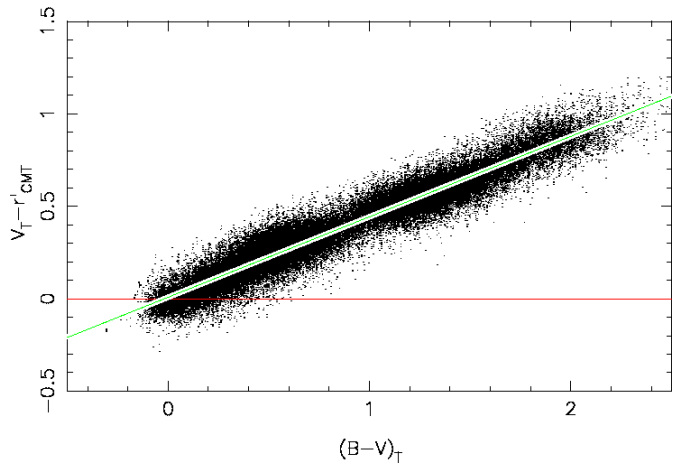}}
\caption{This colour-colour diagram shows the relationship between the
Tycho 2 magnitudes and the CMT instrumental magnitude, $r'_{\rm CMT}$.
Only stars brighter than $\rm V_T<10$ were used in the determination.
The green line is a least-squares linear fit to the data.}
\label{col1}
\end{figure}

\begin{figure}
\resizebox{\hsize}{!}{\includegraphics[angle=270]{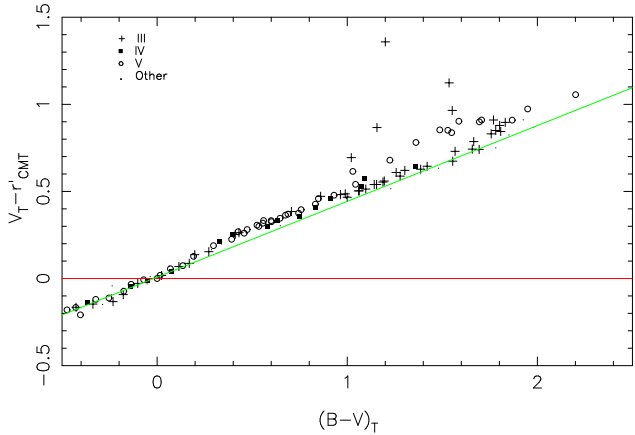}}
\caption{The results of synthetic colour calculations are shown in this plot.
The various symbols stand for different types of star.
The line drawn is the same as that in Figure~\ref{col1}.}
\label{col2}
\end{figure}

Since the CMT only observes in one passband, photometry from the telescope
cannot be placed on a standard photometric system without additional colour
information. However, using the colour data in the Tycho 2 catalogue,
it is possible to calibrate onto the instrumental magnitude system. 
Figure~\ref{col1} shows the least-squares solution used
to determine the linear colour term. This was found to be
$\rm 0.435\times(B-V)_T$.

By using this {\sl a priori} colour term, the $V_T$ of the Tycho 2 standards
can be converted into the natural system of the CCD and filter combination.
This is close to the Sloan $r'$ passband, but will be on the Vega scale rather
than the spectrophotometric AB$_\nu$ magnitude system (\cite{SloanPhot}).

To test that this colour term was reasonable, synthetic colours were
generated in a manner similar to that in \cite{Photometry}. Again, the
difference being the use of data from \cite{Pickles} rather than from
\cite{GS}. The resulting colour-colour diagram is shown in Figure~\ref{col2}
along with a line representing the colour term that was determined from the
CMT data. At the red end, the line lies close to the giant stars rather than
the main sequence stars. This agrees with the expectation that for
V$<$10,
almost all red stars are giants
(Besan\c{c}on Galaxy Model, \cite{BesGalMod}).

When determining the zero point, a weighted least-squares solution is used
along with a rejection filter to identify outliers. A number of diagnostics are
determined in this solution. The main ones are the scatter of
the residuals from the calibration and the magnitude limit. The former gives
an indication of the quality of the photometric conditions, while the latter
will show the presence of cloud. In addition to this, a median filter is
applied to the photometric residuals as a function of time in order to
determine if an exposure was interrupted by cloud. Depending on the level of
the cloud opacity, parts of a data frame can be flagged as non-photometric
or not suitable for the survey. In the latter case, this is when the
effective magnitude limit of a part of an exposure is brighter than 
$r'_{\rm CMT}=16$.
This information is then passed to the selection programme so that another
observation can be rescheduled. Non-photometric observations are accepted
into the survey, since the primary purpose of the survey is astrometry.


\subsection{Linearized photometry scale}\label{APMCAL}

Checks have been carried out using timed exposures to confirm the linearity
of the CCD. These show that the CCD is linear to at least the 1\% level all
the way up to when the CCD saturates at about 60,000 counts. This
corresponds to an $r'_{\rm CMT}$ magnitude of 8--9 for typical observing
conditions.

Since the photometry is derived from isophotal intensities a correction is
required to obtain total magnitudes. Since the image profiles have been
found to be exponential, the appropriate correction from \cite{IrwinHall}
has been applied.
Investigations into an alternative method of linearizing the photometry scale
using a variant of the algorithm developed by
\cite{APMCAL} have also been carried out. 


\subsection{Photometric extinction}\label{extinction}
With the photoelectric micrometer, the photometric solution that was carried
out each night followed a more classical solution (\cite{CMC11}), with the
extinction in V being calculated as part of the photometric solution. This
data was published regularly on the
Internet\footnote{http://www.ast.cam.ac.uk/$\sim$dwe/SRF/camc\_\,extinction.html}
and covered  the years 1984--1998.

From June 1998, since a CCD was being used, the observing strategy and the
part of the sky being observed, prevented a classical solution of the
extinction being made since there was not a large enough range in $\sec z$.
However, by assuming that the zero point of the photometric solution
only changes gradually over time and that the lowest measurable extinction
would correspond to the dust-free value for $r'_{\rm CMT}$ (0.09), 
it is possible to derive an extinction value
for each relatively stable night.

To do this, the zero point determined from the photometry is first corrected
for exposure time ($\equiv\cos\delta$) and then a simple linear model is
applied to account for the change in sensitivity of the system. Finally, a
correction for $\sec z$ is applied in order to produce an extinction value.
Although this is only for the $r'_{\rm CMT}$ passband, it is possible to generate
extinction values for other passbands using the data contained in 
\cite{TN31}.

Since March 1999, extinction values for $r'_{\rm CMT}$ have been published on the
Internet, {\sl cf.} the earlier data. In these tables the mean for each night
is given using only those CCD frames that were considered photometric. On
average, each data frame has 30--40 calibrating stars in it.

Even though this is not a customized
extinction monitor, such as \cite{Hogg}, the CMT extinction data is 
currently the only
source of regular extinction measurement available on the La Palma site and
thus provides a valuable service.


\section{Error estimation}
In order to measure the internal errors of the catalogue, data can be used
from the overlap regions and repeat observations. The results of such an
analysis are given in Table~\ref{TabErrors}. However, it must be understood
that since correlations exist between the data and unaccounted for
systematic errors, these measurements will tend to underestimate the true,
external, errors of the data.

\begin{table}
\caption{The median internal and external errors for the CMT. 
The units for the RA
and declination are milli arc seconds and those for the magnitudes are
millimagnitudes.
}\label{TabErrors}
 \centering
 \begin{tabular}{rrrr}
  \hline
  \multicolumn{4}{c}{Internal}\\
  \hline
  $r'_{\rm CMT}$ & RA & Dec & Mag\\
  \hline
$<$13 &  21 & 21 &  16\\
   14 &  31 & 26 &  30\\
   15 &  55 & 42 &  58\\
   16 & 112 & 91 & 124\\
  \hline
 \end{tabular}
\qquad
 \begin{tabular}{rrrr}
  \hline
  \multicolumn{4}{c}{External}\\
  \hline
  $r'_{\rm CMT}$ & RA & Dec & Mag\\
  \hline
$<$13&  36 & 37 &  25\\
   14&  45 & 40 &  35\\
   15&  68 & 55 &  70\\
   16& 113 & 90 & 170\\
  \hline
 \end{tabular}
\end{table}

\begin{figure}
\resizebox{\hsize}{!}
{\includegraphics[angle=270]
{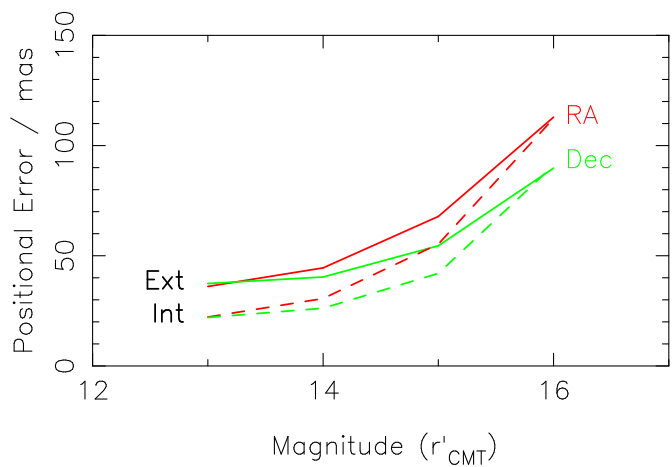}}
\caption{The internal and external positional errors as a function of 
magnitude. The
solid line gives the median external errors and the dashed line shows the
equivalent internal errors. RA is shown in red and declination in 
green. 
}\label{FigErrors}
\end{figure}

The measurement of external errors can be problematical since it requires a
comparison with another catalogue where the errors are either much smaller,
and the residuals yield the external errors directly, or where the errors
are very well determined, and can thus be accounted for in the residuals. In
the former case, not many such catalogues exist and even in those cases the
data is quite sparse, and in the latter, the estimation is dependant on the
external errors of the comparison catalogue being reliable.

A comparison of the CMT data has been carried out with respect to Tycho 2,
but this did not yield very useful results since it was limited to the
brighter end ($r\,^\prime_{\rm CMT}<12$) of the CMT catalogue. 
Also, the
average error of a faint Tycho 2 star, which is in common with the CMT data,
is slightly larger than the average CMT error. In combination with the low
number of stars in these comparisons, this makes a CMT error calculation
difficult to estimate using just Tycho 2 data. Although the results are
quite noisy, it is possible to produce an approximate external error for the
CMT catalogue of 40 mas.

Another technique available involves the use of 2 or more deep comparison
catalogues, where it is then possible to measure the external errors
directly of all the catalogues involved without having to assume any
external error measurements. As with a comparison with a single catalogue,
allowance must be made for proper motions and, if the epoch difference
between the catalogues is large, the errors in the proper motions.

The basis of the technique is the assumption that the 
residuals between any two catalogues result from the quadrature sum of the
external errors. If three catalogues exist, it is possible to derive these
external errors from the three sets of residuals by simple substitution.
If an appreciable epoch difference exists, an allowance must be
made for the errors in the proper motions used in the comparison. This acts
as an additional term in the quadrature sum and has to be removed using the
quoted proper motion error values.

Comparisons with the FASTT and UCAC catalogues (\cite{Stone2},
\cite{Zacharias}) using this technique showed that the astrometric
accuracy of the CMT catalogues, before secondary calibrations (atmospheric
fluctuations and CTE correction) are carried out is 50--80 milli arc seconds
(mas) at the bright end. After such calibrations are applied, the
accuracy improves to 25--45 mas. The dependency of these accuracies as a
function of magnitude is given in Table~\ref{TabErrors}. Although only one
value is quoted per magnitude bin in these tables, there exists a range of
accuracies, as quoted earlier, which is caused by the varying density of
Tycho 2 standards across the sky. For a region of the sky which has more
standards in it, the accuracy of the catalogue at that point will be better.

Figure~\ref{FigErrors} shows these results in graphical form. For bright
stars ($r\,^\prime_{\rm CMT}<13$), the accuracy of the astrometry is about 35 mas for
both RA and declination, but as you go fainter the accuracy in RA becomes
gradually worse than that for declination. The most probable explanation is
that this is caused by further effects resulting from the CTE problem
which the calibration has not yet accounted for. This would only affect RA.

For photometry, the external accuracy estimates are more uncertain since the
comparison catalogues are not primarily photometric ones. Additionally,
there are probably some differences in the passbands used, which would
result in unaccounted colour terms. Assuming the quoted errors from
\cite{Stone2}, a comparison with the FASTT data yields rough estimates for
the external errors and are given 
in the external part of Table~\ref{TabErrors}. The internal photometric
errors were calculated from the overlaps in the same way as that for the
astrometry and are also given in this table. 


\section{Description of the released catalogue}
After the fluctuation calibration has been carried out, the data frames
undergo a further 6-parameter linear fit (see Equation~\ref{Eq6Param}) 
using the Tycho 2 standards. This is done since the residuals have been
reduced and a more accurate fit can be achieved.

The final catalogue consists of the averaged positions from the
calibrated data frames. For the current 
version only a simple algorithm is used whereby
any images within $2''$ of each other are considered as the same
source. However, considering that the average image size is larger than
this, it is likely that this algorithm is sufficient.

The current release of the catalogue (Version 1.0) covers the declination
zone $-3^\circ<\delta<+3^\circ$. The data is available over the
Internet\footnote{Either CDS or http://www.ast.cam.ac.uk/$\sim$dwe/SRF/cmc12/} 
where the
format of the catalogue is described. The main part of the survey extends to
$+30^\circ$ and further releases of the catalogue which will complete the
coverage will
be available in the future from the same location. An extension of the
survey is planned, extending it to $+50^\circ$ in the North and $-15^\circ$
in the South. 



\section{Conclusions}
By upgrading the Carlsberg Meridian Telescope to have a CCD operating in
drift-scan mode, a new lease of life has been breathed into the telescope.
It should be pointed out that this will only be useful over the next ten
years or so. Then, data from astrometric satellites such as DIVA and GAIA
will become generally available and supersede the astrometric accuracy
of what can be achieved from the ground. It is thus important that planned
upgrades of meridian telescopes are carried out as soon as possible and that
the results are published promptly so that the maximum use can be made of
the data.

The results shown here demonstrate that using transfer functions it is possible
to calibrate the fluctuations caused by atmospheric turbulence using just
the Tycho 2 stars. This is a more efficient method than using a subcatalogue
since multiple measurements of the sky are not required.

After this calibration, the external accuracy
achieved for the brightest stars in the survey is 36 mas in right
ascension and declination and 0.025 magnitudes in $r'_{\rm CMT}$ photometry. 

The web site of the telescope is at:\\
http://www.ast.cam.ac.uk/\verb+~+dwe/SRF/camc.html


\begin{acknowledgements}
We thank
Bob Argyle, 
Claus Fabricius, 
Ole Einicke, 
Anton S{\o}rensen,
Jens Klougart.
Niels Michaelsen, 
Torben Knudsen, 
Jose Mui{\~n}os, 
Fernando Beliz{\'o}n 
and 
Miguel Vallejo 
for useful and helpful discussions and advice.

The Institute of Astronomy personnel are part of the Cambridge Astronomical 
Survey Unit which is funded by the Particle Physics \& Astronomy Research
Council of the United Kingdom.


The Danish participation in the project has been funded by the
Copenhagen University Observatory. The Carlsberg Foundation provided
financial support for the initial CCD development.

Thanks are due to Jean-Fran\c{c}ois Le Campion and Michel Rapaport, Bordeaux, 
and Norbert Zacharias, USNO, for early release of data.
\end{acknowledgements}




\appendix

\section{Calibration of CTE feature}\label{secCTE}

Initial comparisons carried out with respect to astrometric standards
published by Stone and co-workers (\cite{Stone1}; \cite{Stone2}) indicated
that a large systematic difference in right ascension existed.

Investigation of the raw data showed that no large image asymmetry was
present. Further comparisons made with data released from Bordeaux confirmed
that the systematic difference was a feature of the CMT data.

It was soon established that this effect was caused by problems with the
charge transfer efficiency of the CCD. Detailed investigations of the CTE
properties of similar CCD chips were carried out by Copenhagen University
Observatory (\cite{cteposterpaper}). Contact with other groups revealed that
this wasn't an isolated case (\cite{Zacharias}). Various solutions are
available to reduce the magnitude of this effect which include fine tuning
the electronics and increasing the temperature of the CCD. However, there is
still a need to calibrate this effect since the data originally taken,
before the final set-up was in place, could still be used and also the 
effect isn't
fully removed by these changes.

\begin{figure}
\resizebox{\hsize}{!}{\includegraphics[angle=270]{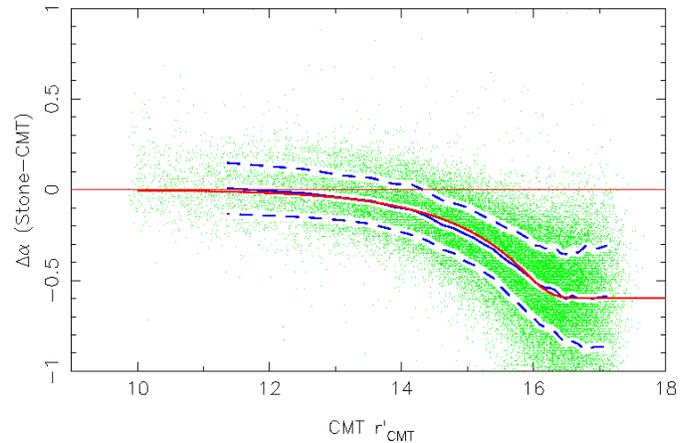}}
\caption{This plot shows the general trend in RA caused by the charge transfer
efficiency problem. This is a comparison, using
observations taken during dark conditions,
with respect to Stone's standards. A three-parameter model has been
fitted to the data.}
\label{cte1}
\end{figure}

The observed effect of this problem is to cause a systematic shift in RA as
a function of magnitude (see Figure~\ref{cte1}). On top of this, the effect
is also a function of the background illumination which reduces the effect
the brighter the sky. This is similar to the reasons needed for a preflash
with early CCDs which also had CTE problems.
A continuous illumination system was considered for the
camera to increase the background level and reduce the effect. However, time
and resources were not available to carry out this experiment.

Initial work compared data from the Stone standard regions and fitted
a 3-parameter function,
\begin{equation}
\Delta\alpha=max(-\exp(ax+b),c),
\end{equation}
where $x$ is the magnitude, for various ranges of background value. A
problem with this function was that no trend could be seen in the parameters
($a,b,c$) as a function of background and thus further examination of the
effect would not be possible.

Additionally, the CTE characteristics changed on 9 August 1999, thus a new
calibration would be needed. This was indicative of a progressive failure of
the camera controller circuitry. It failed totally at the beginning of
October 1999 and had to be replaced.

Since the equator had already been covered by
the survey, it was unclear whether we had enough data, for all periods, to
calibrate the problem properly.
Although brief comparisons had also been carried out with respect to other
telescopes (Bordeaux), there was insufficient data to provide a calibration.
It was thus desirable to compare with other data sets.

\begin{table*}
\begin{tabular}{ccccccr}
Period & Start & End & Controller & Temperature & Number of    & \% \\
       &       &     &            &             & observations & \\
       &       &     &            &             & (1000's)     & \\
0   & 990331 & 990408 & 1 & $-65\degr$C &    518 & 1 \\
1   & 990409 & 990806 & 1 & $-58\degr$C & 10,721 & 23 \\
2.1 & 990809 & 990821 & 1 & $-58\degr$C &    976 & 2 \\
2.2 & 990823 & 990923 & 1 & $-58\degr$C &  1,630 & 3\rlap{.5} \\
2.3 & 990924 & 991007 & 1 & $-58\degr$C &  1,199 & 2\rlap{.5} \\
3   & 991101 & 991213 & 2 & $-65\degr$C &  1,042 & 2 \\
4   & 991214 & 000127 & 2 & $-54\degr$C &  1,741 & 4 \\
5   & 000128 & 000930 & 2 & $-30\degr$C & 28,626 & 62 \\
\end{tabular}
\caption{This table shows the different calibration periods identified
along with some of the characteristics of the periods. The data analysed only
covers up to the end of September 2000.}
\label{table1}
\end{table*}

Palomar Observatory Sky Survey II (POSS2) data was obtained for 27 fields.
These had been scanned by the APM (\cite{APM}). This data was matched with 
the CMT CCD data
and the residuals plotted as a function of magnitude (cf.~Figure~1) for
various sky brightness ranges.
Although this data was less accurate than that of Stone, more data was
available.

Initially, the data was split into 5 periods (1--5), depending on which CCD
controller was in operation and what the CCD temperature was. 
The Controller 1 period was split into two due to the change in CTE
characteristics mentioned above. 

While carrying out the overlap analysis (see later) it was realized that the
first data period had different characteristics, so was split into Periods 0
and 1. This corresponds to a change in the CCD temperature caused by vacuum
problems within the camera. Also, the calibration was very unstable for
Period 2, so it was split into three. Table~\ref{table1} shows the different
calibration periods finally adopted.

In analysing this data, it became clear that the ($a,b,c$) model was not
good enough. The replacement model eventually chosen was a simple scaled
model. In this model the functionality with magnitude was taken directly
from the darkest sky bin of the POSS2 data comparison (the one with the
largest effect) and then scaled to fit the other sky brightness bins. Thus
the parameterization had been reduced from 3 to 1, the scale value, which
could then, in turn, be plotted as a function of sky brightness to see if
further parameterization could be carried out. The scale value was defined
to be equivalent to the RA shift at the faint end (in arc seconds).

\begin{figure}
\resizebox{\hsize}{!}{\includegraphics[angle=270]{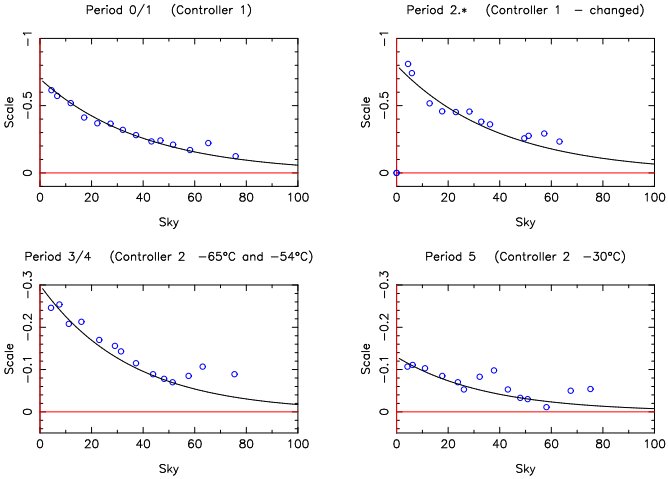}}
\caption{This plot shows the scale value plotted versus sky brightness
for 4 different periods. A simple
exponential has been fitted to the data.}
\label{cte2}
\end{figure}

The results of these calibrations are shown in Figure~\ref{cte2}. A simple
exponential has been fitted to the data:
\begin{equation}scale=z\exp(-sky/y)\label{cte_eqn}\end{equation}
with the data split into four periods: 0/1, 2.{\tt *}, 3/4 and 5.
$y$ and $z$ in this equation are the free parameters of the fit.

These diagrams demonstrate that a simple parameterization exists for the CTE
calibration covering the entire survey. They also clearly show the effect of
the various changes made to the CCD camera. At the start of the project the
maximum size of the CTE shift ($z$) was about $0.7''$. The new controller
reduced this down to $0.3''$ due to improved voltage settings and finally
the increase of the CCD temperature from $-65\degr$C to $-30\degr$C
reduced it further to $\sim0.1''$. 

Work on the UCAC comparisons lead to the implementation of a CTE ``problem
spotter'' in the catalogue generation programme via the overlaps. 
If any of the overlap comparisons
show an RA trend after the CTE calibrations have been carried out, it
indicates a problem in one (or both) of the two frames of the overlap. Many
overlap pairs in the survey were flagged as having problems, thus indicating
that a four-period scaled model, characterized by Figure~\ref{cte2}, was too
simplistic.

An analysis was then carried out on all the residual CTE shifts. Firstly,
all overlaps were determined (14,000 overlaps for 5,600 frames) and the
relative CTE shift for each overlap calculated. At this point we only have
information on the overlaps and do not know the individual contribution from
each frame. Then, for each frame, the median CTE shift for all its overlaps
is calculated, taking care with the signs of these overlap shifts. If
most of the calibrations for the frames are correct, then the median will
give a good approximation to the CTE shift remaining for that frame. We can
then use these frame values to correct the overlap values and then improve
our original estimates of the frame values by iteration. This process
converges after about 5 iterations.

Note that this is all carried out {\em after} a standard CTE calibration
had been applied. Thus, the presence of outliers or increased scatter
in the results indicates a problem with the current calibration.

\begin{figure}
\resizebox{\hsize}{!}{\includegraphics[angle=270]{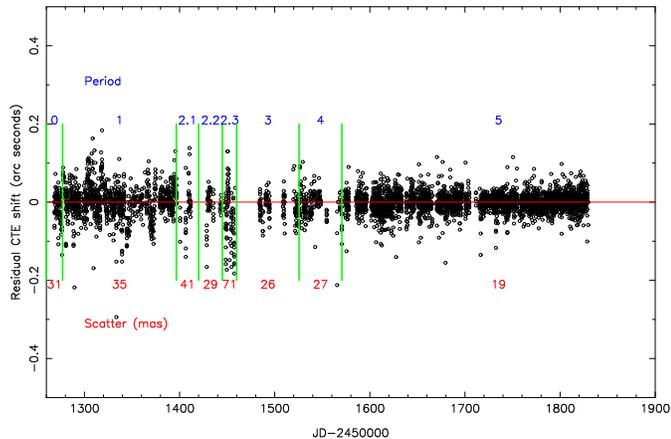}}
\caption{This plot shows the residual CTE shift (after calibration) as a
function of time.}
\label{cte3}
\end{figure}

Figure~\ref{cte3} shows the residual CTE shifts for each frame as a function
of time. Using a figure like this, it was possible to subdivide the original
4 periods, as used for Figure~\ref{cte2}, into the periods indicated in
Table~\ref{table1}. For each period, by plotting the residual against sky
background you can improve the CTE calibration for that period. In addition
to using an exponential model, for some periods, it was necessary to use a
model of the form $scale=a/(b+sky)$, cf. Equation~\ref{cte_eqn}.

Also shown in Figure~\ref{cte3} is the scatter for each period. This gives an
indication as to how well the calibration has been carried out. The
improvement in the calibration as the survey has progressed simply reflects
the reduction in the overall level of the CTE problem.

Period 2 has a high scatter because it generally contains poor data.
Originally, data from this period had a much higher scatter. Closer analysis
showed that a large part of this came from a few nights. Thus, it was
decided to delete 4 nights (5, 6, 15 and 18 September 1999) from the
survey. Doing this loses about 650,000 observations (about 1\% of the data
so far). 

From the scatters shown, it can be seen that the CTE problem has been solved
to about the 100 mas level for the faint end (3 sigma limits), cf. the faint
end random errors of approximately 100 mas.
Figure~\ref{cte1} can be used to gauge how this number translates to brighter
magnitudes. 


\section{Flat-fielding}\label{FlatFielding}

\begin{figure}
\resizebox{\hsize}{!}{\includegraphics[angle=270]{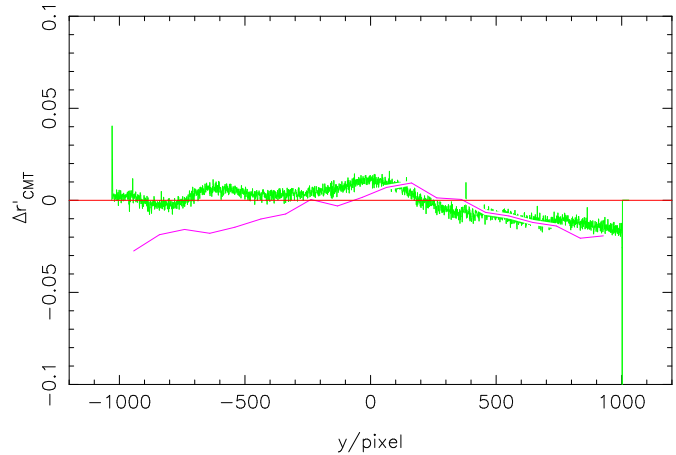}}
\caption{This plot shows a typical flat field for high sky brightness
conditions (noisy solid line). Also shown is the systematic effect in 
magnitude from Figure~\ref{field} (solid line).}
\label{flatplot}
\end{figure}

No multiplicative flat field is applied to the raw frame data in the image
analysis because of the difficulty in separating multiplicative (flat field)
and additive (bias) components and because of the small size (1\%) of the
effect (see Figure~\ref{flatplot}).
Additive (bias and sky) corrections are carried out and the 
image analysis
programme
removes the bias as a one-dimensional function in $y$, combined
with a one-dimensional function in $x$ to remove background variations.

Although the raw image data is not kept, the 
flat-field information is archived every night. Since the CCD camera
takes drift-scan observations, the flat fields are one dimensional.
Not all the frames will have a flat
field since if the sky level is too low then no reliable flat field can
be calculated. About two thirds of the frames do not have any flat-field
information.

Figure~\ref{flatplot} shows a typical flat field from a frame with a high
sky level. Monitoring of the flat fields shows that the main variation seems
to be as a function of sky level. If the sky level is high, then the flat
field is fairly flat ($\pm$1\% level). If the sky level is low, then there 
is about a 10\%
variation at one edge of the CCD, most likely caused by 
just a change in the y-dependent bias level.
Over time there is hardly any
change in the general flat field shape, however, 
occasionally a flat field shows a significant difference of 
a few percent over the  2000 pixels. Closer investigation has shown that
these changes are caused by changes in the background and not the
sensitivity of the CCD.

Various investigations were carried out to determine whether the measured
flat fields represent a multiplicative correction or not due to the problem of
disentangling bias and sky.

One experiment was to apply the flat fields during the initial calibration
after the preliminary data reduction. Normally, each column would be scaled by
the corresponding one in the flat field. For this test, the appropriate
column used was the average $y$ value of an image and the corresponding
scaling factor was applied to the whole image. Since this was already an
approximation, a median filter was applied to the flat field data in order
to reduce the level of noise.

After re-reducing the data, a comparison was made with respect to Tycho 2
This showed that the flat fields were not appropriate
to use since extra systematic effects at the level of $\sim$5\% were
now visible. 
This implies that what is measured as the flat field has at least an additive
component, as pointed out above.

As mentioned before, gradients are seen in some of the flat field
comparisons. These are of order a few percent over the width of the CCD
chip. These tend to happen during periods of bad weather, possibly
indicating partially illuminated cloud causing a real gradient in the
background and hence an additive effect.

To check this, a comparison was carried out 
on one particular frame, which has a 5\%
variation, and a number of overlapping frames. No variation could be seen in
the photometry as a function of $y$ for $r'_{\rm CMT}<14$. This indicated that the
sensitivity was not varying across the chip. A variation {\em is} seen for a
sample of faint images. This is consistent with background levels varying
with $y$ causing problems with the isophotal correction. 

Also shown in Figure~\ref{flatplot} is the magnitude systematic from
Figure~\ref{field}. The obvious feature of this plot
is that one side matches up reasonably well, while the other doesn't. 

Considering the significant differences between the measured flat fields and
the photometric corrections derived from the external standards, it was 
decided that there was no point in using the flat field information in the
photometric calibration.
 

\end{document}